\documentclass[final,5p,times,twocolumn]{elsarticle}
\biboptions{sort&compress}
\usepackage{graphicx}
\usepackage{placeins}

\usepackage{amssymb}

\usepackage[hyperindex,breaklinks]{hyperref} 

\def\MET{\ensuremath{E_{\mathrm{T}}^{\mathrm{miss}}}}
\def\pT{\ensuremath{p_{\mathrm{T}}}}
\def\antibar#1{\ensuremath{#1\bar{#1}}}
\def\ttbar{\antibar{t}}
\def\ET{\ensuremath{E_{\mathrm{T}}}}

\journal{Phyics Letters B}

\begin{document}

\begin{frontmatter}



\title{\vspace{-0.5cm} \flushright{\normalsize{CERN-PH-EP-2012-032}} \\ \center{Search for FCNC single top-quark production at $\sqrt{s}=7$~TeV with the ATLAS detector}}


\author{The ATLAS Collaboration}

\address{}

\begin{abstract}
A search for the production of single top-quarks via flavour-changing neutral-currents is presented.
Data collected with the ATLAS detector at a centre-of-mass energy of $\sqrt{s}=7$~TeV,  
corresponding to an integrated luminosity of 2.05~fb$^{-1}$, are used. 
Candidate events with a semileptonic top-quark decay signature are classified as signal or background-like events
by using several kinematic variables as input to a neural network.
No signal is observed in the neural network output distribution and a Bayesian upper limit is placed on the production cross-section.
The observed upper limit at 95\% confidence level on the cross-section multiplied by the $t \to Wb$ branching fraction is measured to be 
$\sigma_{qg\to t} \times {\cal B} (t\to Wb)<3.9$~pb. This upper limit is converted using a model-independent approach into upper limits on the coupling strengths 
$\kappa_{ugt}/\Lambda < 6.9\cdot 10^{-3}$~TeV$^{-1}$ and $\kappa_{cgt}/\Lambda < 1.6\cdot 10^{-2}$~TeV$^{-1}$, where $\Lambda$ is 
the new physics scale, and on the branching fractions ${\cal B}(t \rightarrow ug) < 5.7\cdot 10^{-5}$ and ${\cal B}(t \rightarrow cg) < 2.7\cdot 10^{-4}$.
\end{abstract}


\end{frontmatter}


\section{Introduction}
\label{sec:intro}

The top quark is the heaviest elementary particle known, with a mass of 
$m_{\rm{top}}=173.2\pm0.9$~GeV~\cite{Lancaster:2011wr} that is close to the electroweak symmetry breaking scale.
For this reason it is 
an excellent object to test the Standard Model (SM) of particle physics. 
The properties of the top quark can be studied from proton-proton ($pp$) collisions at $\sqrt{s}=7$~TeV
with the Large Hadron Collider~(LHC). 
Top-quark pair-production via the strong interaction has been measured at the LHC~\cite{Aad:2011yb,PhysRevD.84.092004}, 
and its cross-section is in good agreement with the prediction of the SM.
Additionally, top quarks can be singly produced through three different 
processes: $t$-channel, $Wt$ associated production, and $s$-channel.
Only $t$-channel single top-quark production has been observed so far~\cite{Aaltonen:2009jj,Abazov:2011rz,Chatrchyan:2011vp}. 
According to the SM of particle physics, flavour-changing 
neutral-current~(FCNC) processes are forbidden at tree level and suppressed at 
higher orders due to the Glashow-Iliopoulos-Maiani mechanism~\cite{Glashow:1970gm}.
Extensions of the SM with new sources of flavour predict 
higher rates for FCNCs involving the top quark; these extensions include new exotic 
quarks~\cite{AguilarSaavedra:2002kr},
new scalars~\cite{Luke:1993cy, Bejar:2000ub}, 
supersymmetry~\cite{Delepine:2004hr, Liu:2004qw, Cao:2007dk,Yang:1997dk},
or technicolour~\cite{Lu:2003yr} 
(for a review see Ref.~\cite{AguilarSaavedra:2004wm}).
If the new particles are heavy, which is consistent with the non-observation of low-mass new particles at the Tevatron and LHC, 
their effects on top-quark FCNCs can 
be parameterised in terms of a set of dimension-six gauge-invariant operators 
\cite{Coimbra:2008qp}. 
The predicted branching 
fractions for top quarks decaying to a quark and a photon, $Z$ boson, or gluon
can be as large as $10^{-5}$ to $10^{-3}$ 
for certain regions of the parameter space in the models mentioned.
For heavy new particles these branching fractions can be large,
if the new particles couple strongly to the SM particles.

According to the corresponding values of the unitary Cabibbo-Kobayashi-Maskawa matrix, the top quark decays 
almost exclusively to a $W$ boson and a $b$ quark. 
FCNC top-quark decays can be studied directly by searching for final 
states with the corresponding decay particles~\cite{Aaltonen:2008aaa,Abazov:2011qf}.
However, the  $t \rightarrow qg$ mode, where $q$ denotes either an up quark $u$ or a charm quark $c$, 
is almost impossible to separate from
generic multijet-production via quantum chromodynamic (QCD) processes, and 
a much better sensitivity can be achieved 
in the search for anomalous single top-quark production.
In the process studied here, a $u$ or $c$ quark and a gluon $g$ coming from the 
colliding protons interact 
to produce a single 
top-quark. The most general effective Lagrangian ${\cal L}_{\rm eff}$ for this process 
resulting from dimension-six operators contains only tensor 
couplings~\cite{AguilarSaavedra:2008zc} and it can be written as~\cite{Hosch:1997gz,Malkawi:1995dm}:
\begin{equation}
{\cal L}_{\rm eff}=g_{s} \sum_{q=u,c} \frac{\kappa_{qgt}}{\Lambda} \bar{t}\sigma^{\mu\nu}T^{a}(f^{L}_q P_L+f^R_q P_R)qG^{a}_{\mu\nu} + h.c. , 
\label{eq:interaction_lagran}
\end{equation}
where the $\kappa_{ugt}$, $\kappa_{cgt}$ are dimensionless parameters that relate the strength of the new coupling to the strong coupling constant $g_{s}$. ${\Lambda}$ 
is the new physics scale, related to the mass cutoff scale above which the effective theory breaks down.
$T^{a}$ are the Gell-Mann matrices~\cite{citeulike:1282207} and $\sigma^{\mu\nu}=\frac{i}{2}[\gamma^{\mu},\gamma^{\nu}]$ transforms as a tensor under the Lorentz group.
The $f^{L,R}_q$ are chiral parameters normalised such that: $|f^{L}_q|^2 + |f^{R}_q|^2 = 1 $. 
The operator $P_L = \frac{1}{2}(1-\gamma^5)$ performs a left-handed projection, while $P_R = \frac{1}{2}(1+\gamma^5)$ performs a right-handed projection, 
where $\gamma^5$ represents the chirality operator. $G^{a}_{\mu\nu}$ is the gauge-field tensor of the gluon and $t$ and $q$ are the fermion fields of the top and 
light quark, respectively.

The existence of FCNC operators allows not only the production
of top quarks via  $qg\to t$, but also the decays $t \rightarrow qg$.
In the allowed region of parameter
space for  $\kappa_{qgt}/\Lambda$ an experimentally 
favourable situation occurs when the FCNC production cross-section for single top-quarks is several picobarns, while the
branching fraction for FCNC decays is very small, and top
quarks can thus be reconstructed in the SM decay mode $t\to Wb$.
The $W$ boson can decay into quark-antiquark pairs ($W\rightarrow q_1 \bar{q_2}$) or a
lepton-neutrino pair ($W\rightarrow\ell\nu$). In this analysis only the decay into a 
lepton-neutrino pair, the leptonic decay, is considered.
Thus the complete process searched for is $qg\to t \to W(\to \ell\nu)b$. 
Selected events
are characterised by an isolated high-energy lepton (electron or muon), 
missing transverse momentum from the neutrino and exactly one jet, 
produced by the hadronisation of the $b$ quark. 
Events with a $W$ boson decaying into a $\tau$ lepton, where the $\tau$ decays into an
electron or a muon are also selected.
The process studied here can be differentiated from SM single top-quark production
because the latter is usually accompanied by additional jets.

This analysis is the first search for FCNCs involving quarks and gluons at the LHC. A search for 
the $2 \to 1$ process  $qg\to t$ 
was performed by CDF~\cite{Aaltonen:2008qr}, while D0 set limits on $\kappa_{ugt}/\Lambda$ and $\kappa_{cgt}/\Lambda$ 
by analysing the $2 \to 2$ processes $q\bar{q} \to t\bar{u}, ug \to tg$, and $gg \to t\bar{u}$ and their $c$ quark analogues~\cite{Abazov:2010qk}.

\section{Data sample and simulation}
\label{sec:data}

The ATLAS detector~\cite{Aad:2008zzm} is built from a set of 
cylindrical subdetectors, which cover almost the full solid angle\footnote{In the right-handed ATLAS coordinate system, the pseudorapidity 
$\eta$ is defined as $\eta=-\ln[\tan(\theta/2)]$, where the polar angle 
$\theta$ is measured with respect to the LHC beamline. The azimuthal angle 
$\phi$ is measured with respect to the $x$-axis, which points towards the 
centre of the LHC ring.  The $z$-axis is parallel to the anti-clockwise beam 
viewed from above. Transverse momentum and energy are defined as $\pT = 
p\sin\theta$ and $\ET = E\sin\theta$, respectively. The $\Delta R$ distance 
is defined as $\Delta R=\sqrt{(\Delta\eta)^2+(\Delta\phi)^2}$.} around the 
interaction point. 

ATLAS is composed of an inner tracking system close to the 
interaction point, surrounded by a superconducting solenoid 
providing a 2~T axial magnetic field, electromagnetic 
and hadronic calorimeters, and a muon spectrometer.
The electromagnetic calorimeter is a high-granularity liquid-argon (LAr) sampling calorimeter with lead absorber.
An iron-scintillator tile calorimeter
provides hadronic energy measurements in the central pseudorapidity range.
The endcap and forward regions are instrumented with LAr calorimeters for
both electromagnetic and hadronic energy measurements.
The muon spectrometer consists of three large superconducting toroids, a system of trigger chambers,
and precision tracking chambers.

This analysis is performed using $\sqrt{s}=7$~TeV
$pp$-collision data recorded by ATLAS between March 22 and August 22, 2011.
Only the periods in which all the subdetectors were operational are considered, resulting in a
data sample with a total integrated luminosity of $2.05\pm0.08$~fb$^{-1}$~\cite{Aad:2011dr,ATLAS-CONF-2011-116}.

Detector and trigger 
simulations are performed with the standard 
simulation of ATLAS within the {\textsc GEANT4}~\cite{Agostinelli:2002hh,Aad:2010wqa} framework.
The same offline reconstruction methods 
used with data events are applied to the simulated samples. 
Minimum bias events generated by {\textsc PYTHIA}~\cite{Sjostrand:2006za} are used to 
simulate multiple $pp$ interactions, corresponding 
to the LHC operation with 50~ns bunch separation and an average of six 
additional $pp$ interactions per bunch crossing.

For the simulation of FCNC production of single top-quarks, {\textsc PROTOS}~\cite{AguilarSaavedra:2010rx} is used.
The top quarks decay as expected in the SM, 
and only the leptonic decay of the $W$ boson is considered. 
$W$ bosons decaying into a $\tau$ lepton, where the $\tau$ decays into an
electron or a muon are included in both the signal and all background samples.
The \textsc{CTEQ6}~\cite{Pumplin:2002vw} leading-order~(LO) parton distribution functions~(PDFs) are used and 
the hadronisation of signal events is simulated with \textsc{PYTHIA} using the 
AMBT1 tunes~\cite{ATLAS-CONF-2010-031} to the ATLAS collision data.
It has been verified that the kinematics of the signal process are independent of the a priori 
unknown FCNC coupling.

Several SM processes are expected to have the same final-state topology as 
the signal.  
Samples of simulated events for the $t$-channel and $Wt$ single top-quark processes are
generated by the \textsc{AcerMC} program~\cite{Kersevan:2004yg} with the \textsc{CTEQ6} LO PDFs and hadronised with \textsc{PYTHIA};
for the $s$-channel process, the \textsc{MC@NLO}~\cite{FRI-0201} generator with the \textsc{CTEQ6.6}~\cite{Nadolsky:2008zw} PDFs
interfaced to \textsc{HERWIG}~\cite{COR-0001} and \textsc{JIMMY}~\cite{JButterworth:1996zw}.

The \textsc{ALPGEN}~\cite{Mangano:2002ea} program with the \textsc{CTEQ6} LO PDFs is interfaced to \textsc{HERWIG} and \textsc{JIMMY} 
to generate $W+$jets, $Wb\bar{b}$, $Wc\bar{c}$, $Wc$ and $Z+$jets events with up to five additional partons.
To remove overlaps between the $n$ and $n+1$ parton samples the MLM matching scheme~\cite{Mangano:2002ea} is used.
The double counting between the inclusive $W+n$ parton samples and samples with associated heavy-quark pair-production is removed utilising an overlap-removal method based on $\Delta R$ matching.
The parameters of \textsc{HERWIG}, with the \textsc{MRST LO**}~\cite{Sherstnev:2007nd} PDFs,  and \textsc{JIMMY} are tuned to ATLAS collision data with the corresponding AUET1 tunes~\cite{PUB-2010-014}.
Diboson backgrounds from $WW$, $WZ$ and $ZZ$ events are simulated using \textsc{HERWIG}.
For the generation of SM $t\bar{t}$ events the \textsc{MC@NLO} generator with the \textsc{CTEQ6.6} PDFs is used. 
The parton shower and the underlying event are added using \textsc{HERWIG} and \textsc{JIMMY}.

\section{Event selection}
\label{sec:evtsel}
Events are considered only if they were accepted by a single-lepton trigger~\cite{Aad:2011xs}.
The single-muon trigger threshold was  $\pT = 18$ GeV,
and the single-electron trigger threshold was raised from an $E_{\rm T}$ of 20 GeV to 22 GeV for higher LHC luminosities.

Electron candidates are defined as clusters of cells in the electromagnetic calorimeter associated 
with a well-measured track fulfilling several quality requirements~\cite{Aad:2011mk}.
Electron candidates are required to satisfy \pT\ $> 25$ GeV and 
$\mathrm{|\eta_{\rm clus}| <}$ 2.47, where $\mathrm{\eta_{\rm clus}}$ is the
pseudorapidity of the cluster of energy deposits in the calorimeter. 
A veto is placed on 
candidates in the calorimeter barrel-endcap transition region, 
$1.37 < |\eta_{\rm clus}| < 1.52$, where 
there is limited calorimeter instrumentation.
High-\pT~electrons associated with the $W$-boson decay can be mimicked by hadronic jets reconstructed as electrons,
electrons from decays of heavy quarks, and photon conversions.  
Since signal electrons from the $W$-boson decay are typically isolated from hadronic jet activity, these backgrounds can be suppressed via isolation 
criteria which require minimal calorimeter activity and only low track $\pT$ in an $\eta$-$\phi$ cone around the electron candidate.
Calorimeter isolation requires the sum of the $E_{\rm T}$ in cells within a cone of $\Delta R = 0.3$ around each electron with $ \pT > 25$ GeV to
satisfy $\sum E_{\rm T}(\Delta R < 0.3)/\pT < 0.15$. Similarly, the scalar sum of the $\pT$ of tracks around the electron must satisfy
$\sum \pT(\Delta R < 0.3)/\pT < 0.15$. The electron track $\pT$ and the $E_{\rm T}$ in associated cells are excluded from $\sum \pT(\Delta R < 0.3)$ and
$\sum E_{\rm T}(\Delta R <0.3)$, respectively.
Muon candidates are reconstructed by matching track segments or complete tracks in the muon spectrometer with the inner detector tracks.
The final candidates are required to have a transverse momentum 
\pT $> 25$ GeV and to be in the pseudorapidity region of $|\eta|<2.5$.
Isolation criteria are applied to reduce background events in which a high-$\pT$ muon is produced
in the decay of a heavy quark.  For the transverse energy within a cone of $\Delta R=0.3$ about the muon direction, 
$\sum E_{\rm T}(\Delta R < 0.3)/\pT < 0.15$ is required, 
while the scalar sum of transverse momenta of additional tracks inside a $\Delta R=0.3$ cone around the muon 
must satisfy $\sum p_{\rm T}(\Delta R < 0.3)/\pT < 0.10$.  
Candidate events are required to have exactly one isolated lepton ($\ell$).

Jets are reconstructed using the anti-$k_{t}$ algorithm~\cite{Salam:2007xv} with 
the distance parameter $R$ set to 0.4.
The jets are then corrected from the raw calorimeter response to the energies of the reconstructed particles
using $\pT$- and $\eta$-dependent factors,
derived from simulated events and validated with data~\cite{Aad:2011he}.
Since the signal process  gives rise to only one high-\pT\ jet,
exactly one  reconstructed jet with $\pT>25$~GeV is required. 

The magnitude of the missing transverse momentum $\MET$ is defined as $\MET = | \vec{E}_{\mathrm{T}}^{\mathrm{miss}} |$, 
where $\vec{E}_{\mathrm{T}}^{\mathrm{miss}}$ is calculated using 
the calibrated three-dimensional calorimeter energy clusters associated with the jet 
together with either the calibrated calorimeter energy cluster associated with an electron or the $\pT$ of a muon track~\cite{Aad:2012re}.
Transverse energy deposited in calorimeter cells but not associated with any high-$\pT$ object is also included
in the $\MET$ calculation. Due to the presence of a neutrino 
in the final state of the signal process, $\MET > 25$~GeV is required. 
To further reduce the number of multijet background events, which are characterised by low $\MET$ and 
low values of reconstructed $W$-boson transverse mass 
$m_{\mathrm{T}}^{W}~=~\sqrt{ 2 \left[p_{\rm T}^{\rm lep} \MET  - \vec{p}_{\rm T}^{\rm lep} \cdot \vec{E}_{\mathrm{T}}^{\mathrm{miss}}\right]}$,
the event selection requires $m_{\mathrm{T}}^{W} + \MET >$ 60 GeV.

Finally, the selected jet has to be  
identified ($b$-tagged) as a $b$-quark jet. The tagging algorithm exploits the properties 
of a $b$-quark decay in a jet using neural-network techniques and   
the reconstruction of a secondary vertex, and has an identification efficiency measured to be about 57\% 
in $\ttbar$ events~\cite{ATLAS-CONF-2011-102}. Only 0.2\% of light-quark jets and 10\% of $c$-quark jets are mis-tagged as $b$-quark jets.
The following samples are defined for this analysis: a ``$b$-tagged sample'' with exactly 
one $b$-tagged jet, and a ``pretagged sample'' without any $b$-tagging requirement.

Assuming a cross-section of 1~pb for FCNC single top-quark production, about 113 signal events in 2.05 fb$^{-1}$
of collision data are expected in the $b$-tagged sample.

\begin{figure*}[!t]
\begin{center}
\includegraphics[width=0.32\textwidth]{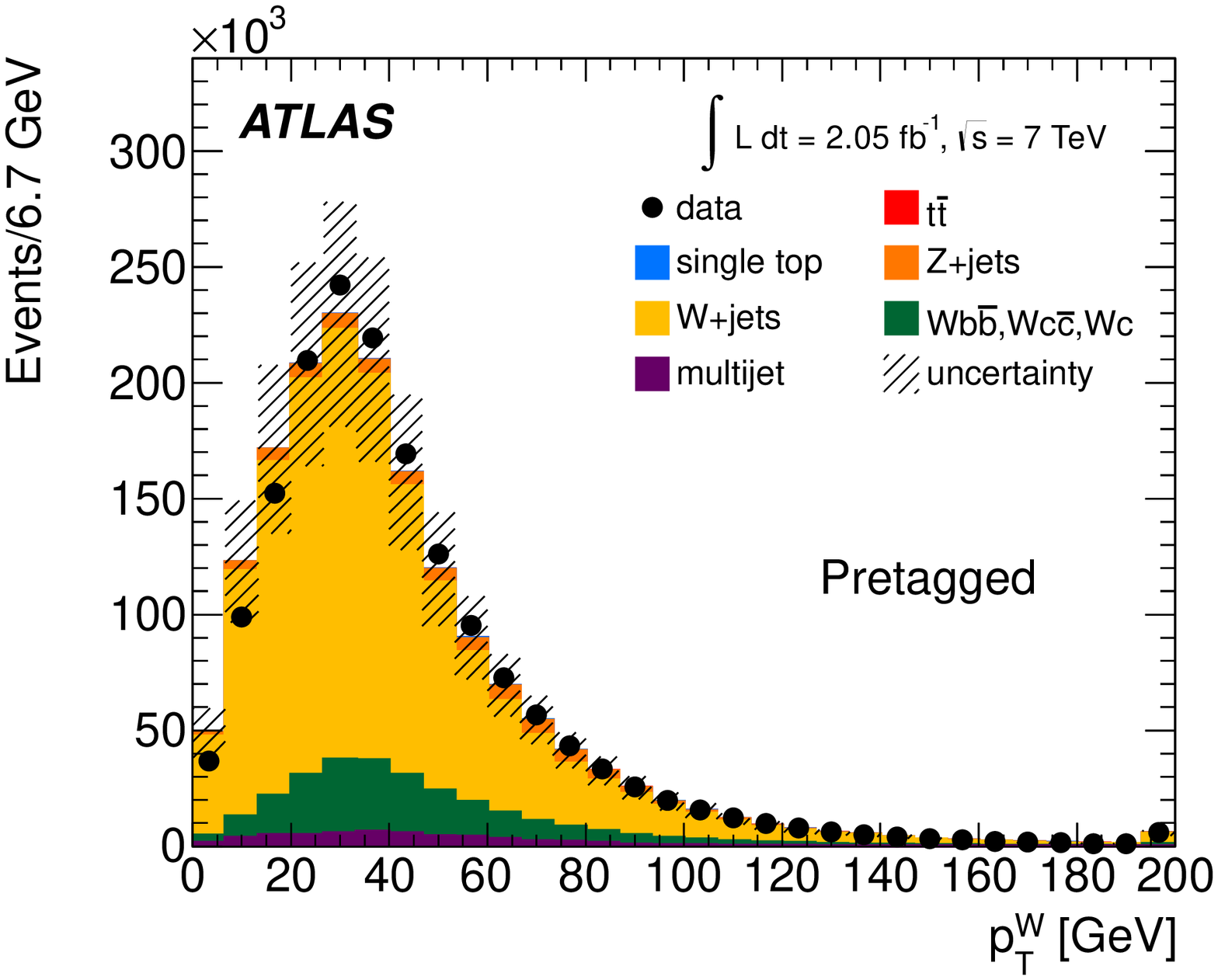}
\includegraphics[width=0.32\textwidth]{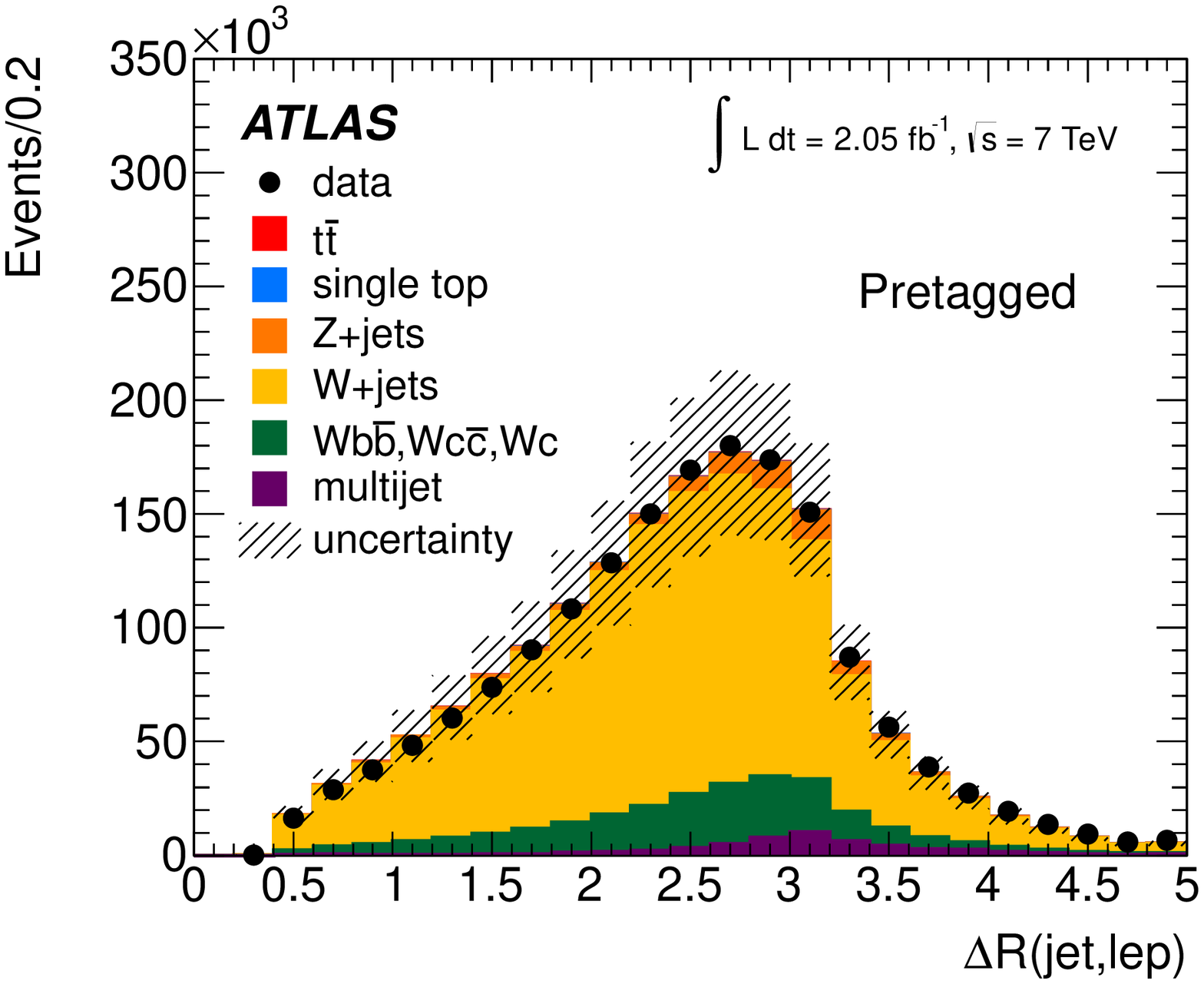}
\includegraphics[width=0.32\textwidth]{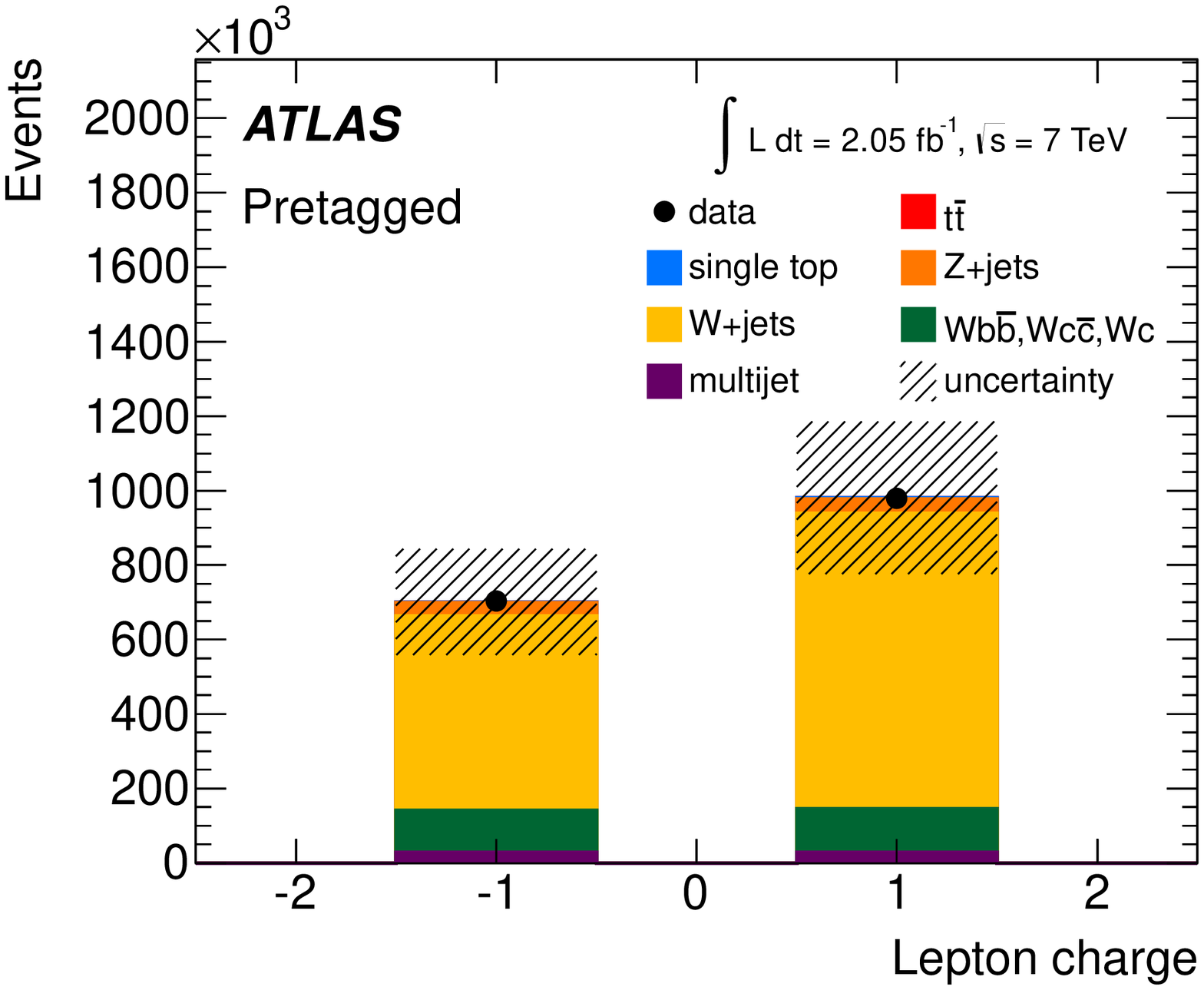}\\[-1em]
(a)\hspace*{0.32\textwidth} (b) \hspace*{0.32\textwidth} (c)\\
\includegraphics[width=0.32\textwidth]{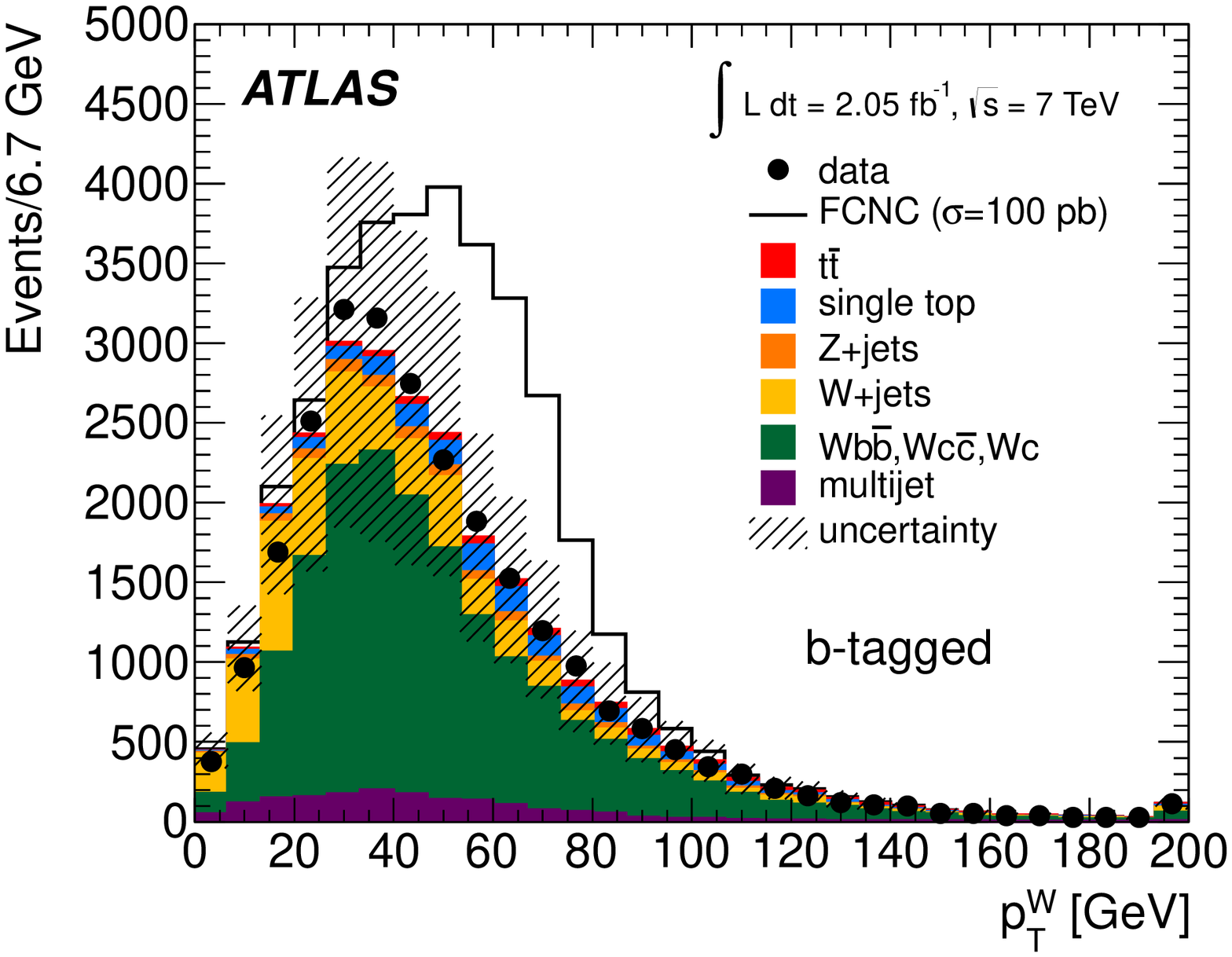}
\includegraphics[width=0.32\textwidth]{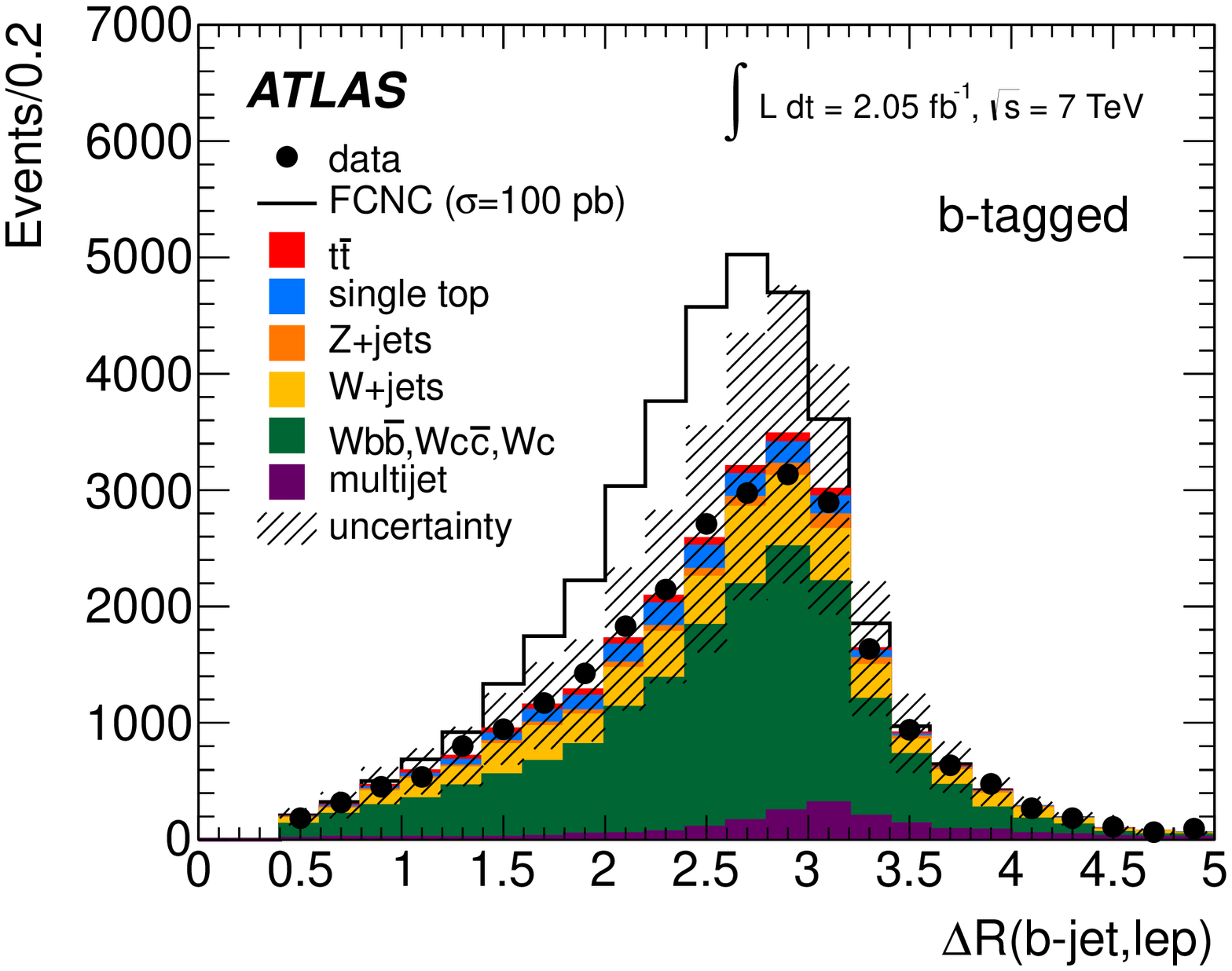}
\includegraphics[width=0.32\textwidth]{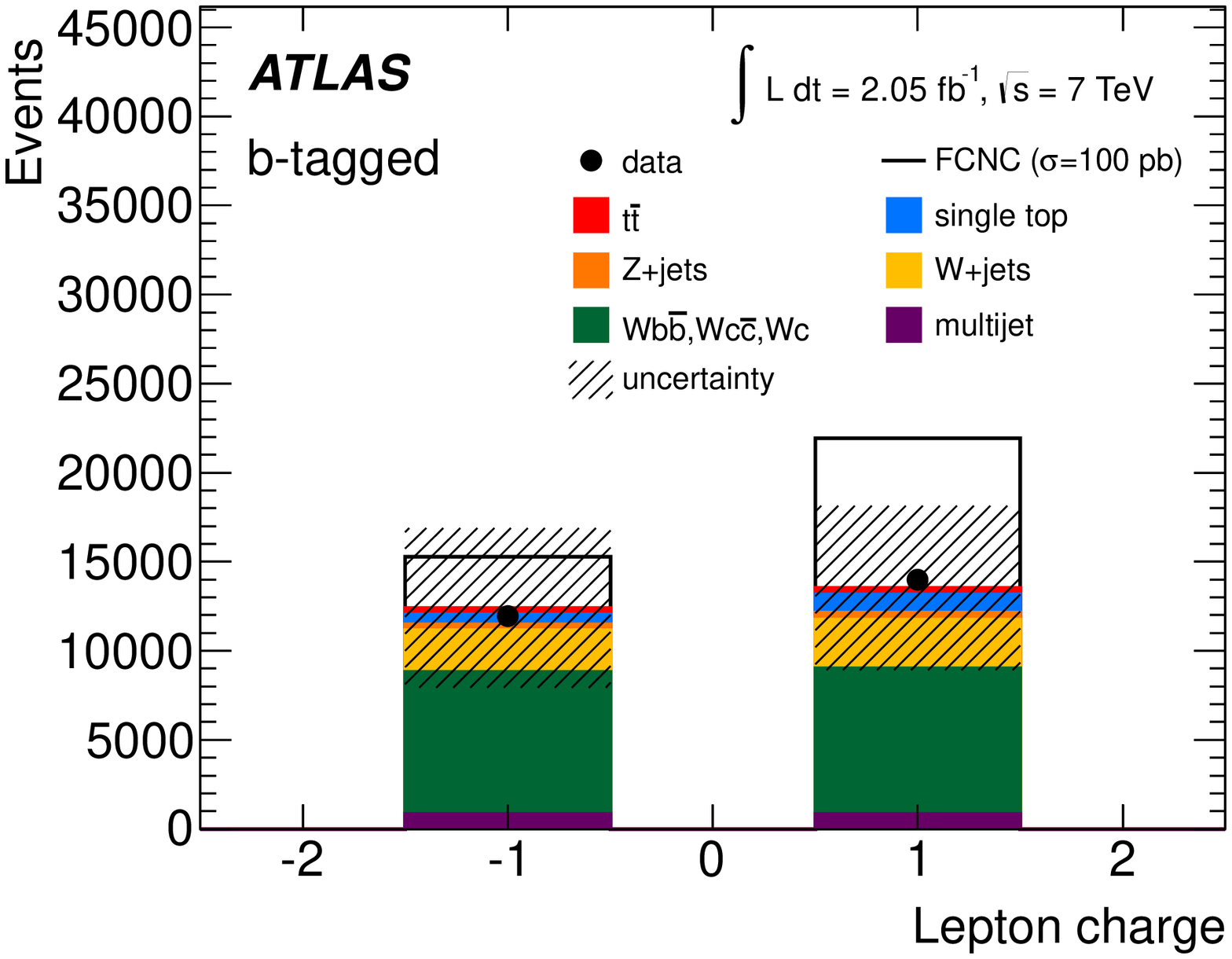}\\[-1em]
(d)\hspace*{0.32\textwidth} (e) \hspace*{0.32\textwidth} (f)\\
\caption{\label{fig:bestthree}
Kinematic distributions of the three most significant variables normalised to the number of observed events 
for the pretagged selection (top) and in the $b$-tagged selection (bottom), for the 
electron and muon channel combined: (a), (d) transverse momentum of the $W$ 
boson, (b), (e) $\Delta R$ between the jet and the lepton and (c), (f) charge of 
the lepton. In these distributions the signal contribution is shown stacked on top of the backgrounds, 
with a normalisation corresponding to a cross-section of 100 pb. 
The hatched band indicates the statistical uncertainty from the sizes of the simulated samples and the uncertainty in the background normalisation.}
\end{center}
\end{figure*}

The normalisations for the various background processes are estimated either by using the experimental data
or by using Monte Carlo simulation scaled to the theoretical cross-section predictions.
For the $W$+jets and $Z$+jets backgrounds the kinematic distributions are modelled using simulated events, while the inclusive cross-sections  
are calculated to next-to-next-to-leading order~(NNLO) with \textsc{FEWZ}~\cite{Anastasiou:2003ds}.
The dominant $W$+jets background process is $Wc$ production, whose $k$-factor is obtained by comparing the NLO and LO cross-sections calculated using
\textsc{MCFM}~\cite{Campbell1999}.
The $W$+(1 jet) and $Z$+(1 jet) background normalisation uncertainties are estimated from the uncertainty in the cross-section of the 
$W/Z$+(0 jet) process and the uncertainty in the cross-section ratio of $W/Z$+(1 jet) to $W/Z$+(0 jet).
A cross-section uncertainty of 4\% is assigned for the $W/Z$+(0 jet) process. Variations consistent with experimental data are made
in \textsc{ALPGEN} to the factorisation and normalisation scale and to the matching parameters, and yield a 24\% 
uncertainty on the cross-section ratio.
Background contributions from the heavy-quark processes $Wb\bar{b}$, $Wc\bar{c}$ and $Wc$ have relative uncertainties of 50\%, estimated using a tag-counting method in control regions.
The \ttbar\ cross-section is normalised to the approximate NNLO-predicted value obtained using \textsc{HATHOR}~\cite{Aliev:2010}. The SM single 
top-quark production cross-section is also calculated to approximate NNLO~\cite{Kidonakis:2011wy,Kidonakis:2010ux,Kidonakis:2010tc}.
A theoretical uncertainty of 10\% is assigned for SM top-quark production.
The normalisation of the cross-section for production of diboson events is
 obtained using NLO cross-section predictions and has an uncertainty of 5\%.

Multijet events may be selected if a jet is misidentified as an isolated lepton or 
if the event has a non-prompt lepton that appears isolated.
A binned maximum-likelihood fit to the $\MET$ distribution is used to estimate the multijet background normalisation. 
A template of the multijet background is modelled using electron-like jets selected from
jet-triggered collision data and is referred to as a jet-electron model. Each jet has to fulfil the same $p_{\rm T}$ and
$\eta$ requirements as a signal lepton, contain at least four tracks to reduce 
the contribution from converted photons, and deposit 80--95\% of its energy in
the electromagnetic calorimeter. 
The uncertainty in the multijet background normalisation is estimated to be 50\% by fitting the distribution of $m_{\mathrm{T}}^{W}$
instead of $\MET$, and using jet-electron models built from jet-triggered data samples with different average numbers of inelastic $pp$
interactions per event. The shape of the jet-electron data sample is used to model the multijet background shape in the electron and muon channels.
The validity of the model in both channels is verified by comparing distributions of multijet-sensitive variables to observed data. 

In the $b$-tagged sample 26223 events are observed in data compared to a prediction
of $24000 \pm 7000$ events from our estimates of SM backgrounds. Table~\ref{tab:sel_tag} summarises the event yield for each of the background
processes considered. Each event yield uncertainty in Table~\ref{tab:sel_tag} combines the statistical uncertainty, 
originating from the limited size of the used samples, with the uncertainty in the cross-section or normalisation.
\begin{table}[h]
  \centering
  \caption{Number of observed data events and expected number of background events for the $b$-tagged sample. 
           The uncertainties include the statistical uncertainty from the size of the simulated sample and 
           the uncertainties on the cross-section and the multijet normalisation.}
  \label{tab:sel_tag}
\begin{tabular}{l|rcr} 
\hline
\hline
 Process     &   \multicolumn{3}{c}{Expected events} \\
\hline  
SM single top& 1460 & $\pm$ & 150 \\
$t\bar t$&  660 & $\pm$ & 70 \\
$W$+light jets &4700 & $\pm$ & 1100 \\
$Wb\bar{b}$/$Wc\bar{c}$+jets& 2700 & $\pm$ & 1500 \\
$Wc$ + jets & 12100 & $\pm$ & 6700 \\
$Z$+jets/diboson & 700 & $\pm$ & 170 \\
Multijet & 1600 & $\pm$ & 800 \\
\hline
Total background  & 24000 & $\pm$ &7000 \\
\hline
Observed & \multicolumn{3}{c}{26223} \\
\hline
\hline
\end{tabular}
\end{table}

\section{Data analysis}
Given the large uncertainty in the expected background and the small number of expected signal events estimated in Section~\ref{sec:evtsel}, 
multivariate analysis techniques are used to separate signal events from background events.
We use a neural-network classifier~\cite{Feindt:2006pm} that combines a 
three-layer feed-forward neural network with a complex robust 
preprocessing. 
In order to improve the performance and to avoid overtraining, Bayesian regularisation~\cite{MacKay92b} is
implemented during the training process. 
The network infrastructure 
consists of one input node for each of the 11 input variables plus one bias node, 13 nodes in the hidden layer,
and one output node which gives a continuous output in the interval [$-1,1$].
The training is done with a mixture of 50\% signal and 50\% background events using about 650000 events, 
where the different background processes are weighted according to their expected numbers of events.

The $qg\to t\to b\ell\nu$ process is characterised by three main differences from SM processes that pass the event selection cuts.
Firstly, in single top-quark production via FCNCs, 
the top quark is produced almost without transverse momentum. Therefore the $p_{\rm T}$ distribution of the top quark is much softer 
than the $p_{\rm T}$ distribution of top quarks produced through SM top-quark production,
and the $W$ boson and $b$ quark from the top-quark decay are almost back-to-back with an opening angle near $\pi$.
Secondly, unlike in the $W/Z$+jet and diboson backgrounds, the $W$ boson from the top-quark decay has a very high momentum and
its highly-boosted decay products have small opening angles.
Lastly, the top-quark charge asymmetry differs between FCNC processes and SM processes. The FCNC processes are
predicted to produce four times more single top quarks than anti-top quarks, whereas in SM single top-quark production and all other SM
backgrounds this ratio is at most two.
All possible discriminating variables such as momenta, relative angles, pseudorapidity, reconstructed particles masses, and lepton electric charge were explored,
including variables obtained from the reconstructed $W$ boson and the top quark.
To reconstruct the four-momentum of the $W$ boson, the neutrino four-momentum 
is derived from the measured $\vec{E}_{\mathrm{T}}^{\mathrm{miss}}$ 
since it cannot be measured directly. The neutrino longitudinal momentum, $p_{z}^{\nu}$, is calculated by 
imposing a kinematic constraint on the $m_{W}$ invariant mass.
The twofold ambiguity is resolved by choosing the smallest $|p_{z}^{\nu}|$ solution, since the $W$ boson is expected to be produced with small pseudorapidity.
The top-quark candidate is reconstructed by adding the four-momentum of the $b$-tagged jet to the four-momentum of the reconstructed $W$ boson.

Eleven variables were selected as input to the neural network after testing for each variable the agreement between the background model and observed events
in both the large sample of pretagged events and the $b$-tagged sample.
The first ten variables are the charge and the $p_{\mathrm{T}}$ of the lepton, the $p_{\mathrm{T}}$, $\eta$ and mass of the $b$-tagged jet, 
the $\Delta R$ between the $b$-tagged jet and the charged lepton, the $\Delta R$ between the $b$-tagged jet and the reconstructed $W$ boson,
the opening angle $\Delta\phi$ between the directions of the $b$-tagged jet and the reconstructed $W$ boson, the $p_{\mathrm{T}}$ 
of the $W$ boson and the reconstructed top-quark mass.
The last variable considered in the neural network is the $W$-boson helicity.
This is calculated as $\cos\theta^{*}$, the cosine of the angle between the momentum
of the charged lepton in the $W$-boson rest-frame and the momentum of the $W$ boson as
seen in the top-quark rest-frame. 
Table~\ref{trainingVars} shows a summary of the used variables ordered by their importance.
The importance of the variables is estimated using an iterative procedure,
removing one variable at a time and recalculating the separation power. 
The ordering is done in terms of relevance defined as standard deviations of the additional separation power given by each variable.
Distributions of the three most important variables in the pretagged sample and the $b$-tagged sample, normalised to the number of observed events,
are shown in Figure~\ref{fig:bestthree}. Since the neural network benefits from the correlation between variables and is trained
to separate the signal process from all background processes, the naively expected variables are not the most important ones, 
but variables, which are highly correlated to them.
\begin{figure*}[!t]
\centering
\includegraphics[width=0.45\textwidth]{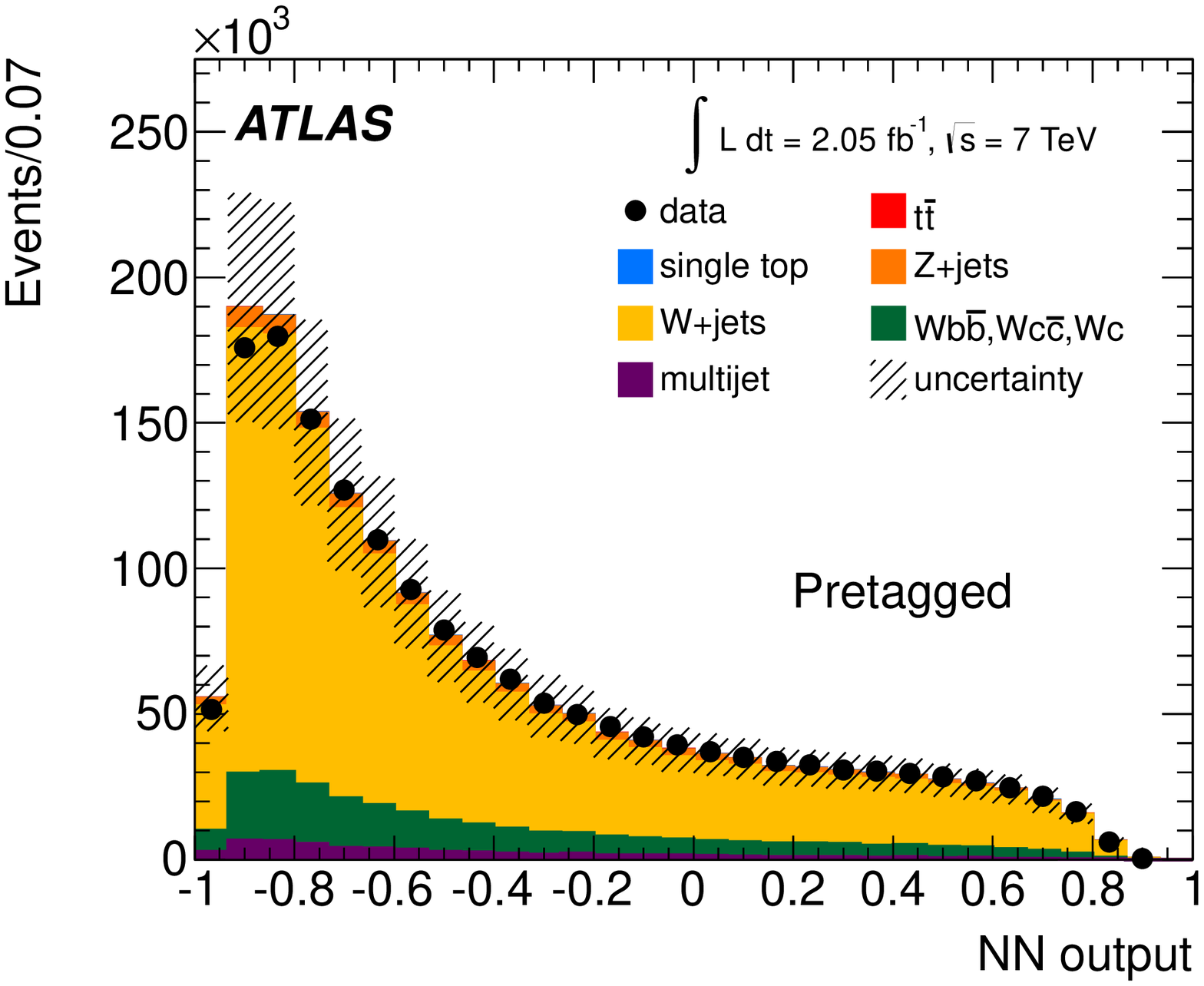}
\includegraphics[width=0.45\textwidth]{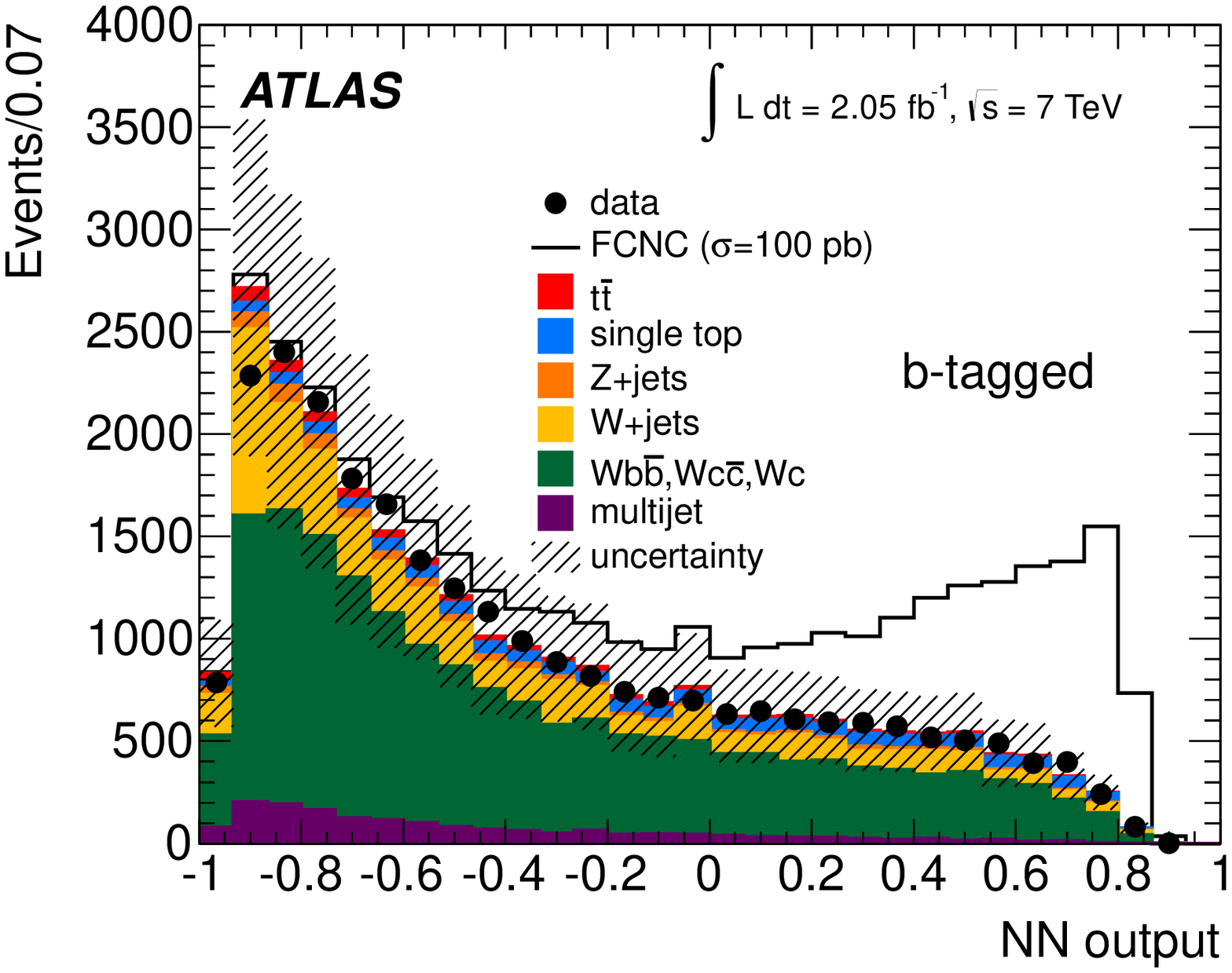}\\[-1em]
(a)\hspace*{0.45\textwidth} (b)
\caption{\label{fig:nn}
(a) Neural network output distribution scaled to the number of observed events in the pretagged sample.
(b) Neural network output distribution scaled to the number of observed events in the $b$-tagged sample.
In these distributions the signal contribution is shown stacked on top of the backgrounds. 
The hatched band indicates the statistical uncertainty from the sizes of the simulated samples and the uncertainty in the background normalisation.}
\end{figure*}
\begin{table}[ht]
\begin{center}
\renewcommand{\arraystretch}{1.15}
  \caption{
  Variables used as input to the neural network ordered by their importance.}
  \label{trainingVars}
\begin{tabular}{lr}
\hline\hline
Variable & Significance ($\sigma$)\\
\hline
$p_{\rm T}^{W}$  & 57 \\
$\Delta R$($b$-jet,lep) & 28\\
Lepton charge & 22\\
$m_{\rm top}$ & 20\\
$m_{b\mbox{-}\rm{jet}}$ & 15\\
$\eta_{b\mbox{-}\rm{jet}}$  & 12\\
$\Delta\phi$($W$,$b$-jet) & 11\\
$p_{\rm T}^{\rm lep}$  & 12\\
$p_{\rm T}^{b\mbox{-}\rm{jet}}$  & 6.5\\
$\cos\theta^{*}$ & 5.7\\
$\Delta R$($W$,$b$-jet) & 5.0\\ 
\hline\hline
\end{tabular}
\end{center}
\end{table}

The resulting neural network output distributions for the various processes, scaled to the number of observed events in the pretagged sample
are shown in Figure~\ref{fig:nn}(a). Figure~\ref{fig:nn}(b) shows these distributions in the $b$-tagged sample. 
Signal-like events have output values close to 1, whereas background-like events are accumulated near $-1$.  
We find good agreement between the neural network output distributions for data and simulated events in 
both the pretagged and $b$-tagged samples.
\section{Systematic Uncertainties}
\label{sec:sys}

Systematic uncertainties affect the signal acceptance, the normalisation of the individual backgrounds, 
and the shape of the neural network output distributions. All uncertainties described below 
lead to uncertainties in the rate estimation as well as distortions of the neural network output distribution 
and are implemented as such in the statistical analysis.

The momentum scale and resolution, as well as the trigger and identification efficiency for 
single leptons is measured in collision data using $Z\rightarrow e e$, $Z\rightarrow\mu\mu$, 
and $W\rightarrow e\nu$ decays and corrective scale factors are applied to the simulation.
Uncertainties on these factors as functions of the lepton kinematics are around 5\%.
To evaluate the effect of momentum scale uncertainties, the event
selection is repeated with the lepton momentum varied up and down 
by the uncertainty. For the momentum resolution uncertainties, the
event selection is repeated with the lepton momentum smeared.
The uncertainty in the jet energy scale, derived
using information from test-beam data, collision data, and
simulation varies between 2.5\% and 8\% (3.5\% and
14\%) in the central (forward) region, depending on jet $p_{\mathrm{T}}$ and
$\eta$~\cite{Aad:2011he}. This includes uncertainties 
due to different compositions of jets initiated by gluons or light quarks in the samples and mis-measurements due to 
close-by jets. Additional uncertainties due to multiple $pp$ interactions
are as large as 5\% (7\%) in the central (forward) region.
Here, the central region is defined  as $|\eta| < 0.8$.
An additional jet energy scale uncertainty of up to 2.5\%, 
depending on the $p_{\mathrm{T}}$ of the jet, is applied for $b$-quark jets 
due to differences between jets initiated by gluons or light quarks as opposed to jets containing $b$-hadrons.
To evaluate the effect of these uncertainties the energy of 
each jet is scaled up or down by the uncertainty and the change is also propagated to the 
missing transverse momentum calculation.
An uncertainty of 2\% is assigned for the jet reconstruction
efficiency based on the agreement between
efficiencies measured in minimum bias and QCD dijet events
and simulated events~\cite{ATLAS-CONF-2010-054}.
For the $b$-tagging efficiencies and mis-tag rates, 
jet $p_{\mathrm{T}}$- and $\eta$-dependent scale factors 
are applied to match simulated distributions
with observed distributions and have uncertainties
from 8--16\% and 23--45\%, respectively~\cite{ATLAS-CONF-2011-102}.

Systematic effects from mis-modelling in event generators
are estimated by comparing different generators and varying parameters for the event generation.
The effect of parton shower and hadronisation modelling uncertainties 
is evaluated by comparing two {\sc AcerMC} samples interfaced
to {\sc HERWIG} and {\sc PYTHIA}, respectively.
The amount of initial and final state radiation is
varied by modifying parameters in {\sc PYTHIA}.
The parameters are varied in a range comparable
to those used in the Perugia Soft/Hard tune
variations~\cite{Skands:2010ak}.
These uncertainties, the parton shower modelling and variations of initial and final state radiation
are evaluated for all processes involving top quarks including the signal.
The impact of the choice of PDFs in the simulation is studied by re-weighting the events
according to PDF uncertainty eigenvector sets (CTEQ6.6, MSTW2008~\cite{Martin:2009iq})
and estimated following the procedure described in~\cite{Campbell:2006wx}.
The uncertainties for the two PDF sets are added in quadrature. 
To account for uncertainties connected with the simulation of the $W$+jets sample several parameters in 
the generation of these samples are varied and event kinematics are compared.
The uncertainty in the measured integrated luminosity
is estimated to be 3.7\%.

The dominant uncertainties are the uncertainties in the jet energy scale, 
the initial and final state radiation variations, and uncertainties
in the $b$-tagging efficiencies and mis-tag rates.

\begin{figure}[!h]
\centering
\includegraphics[width=0.45\textwidth]{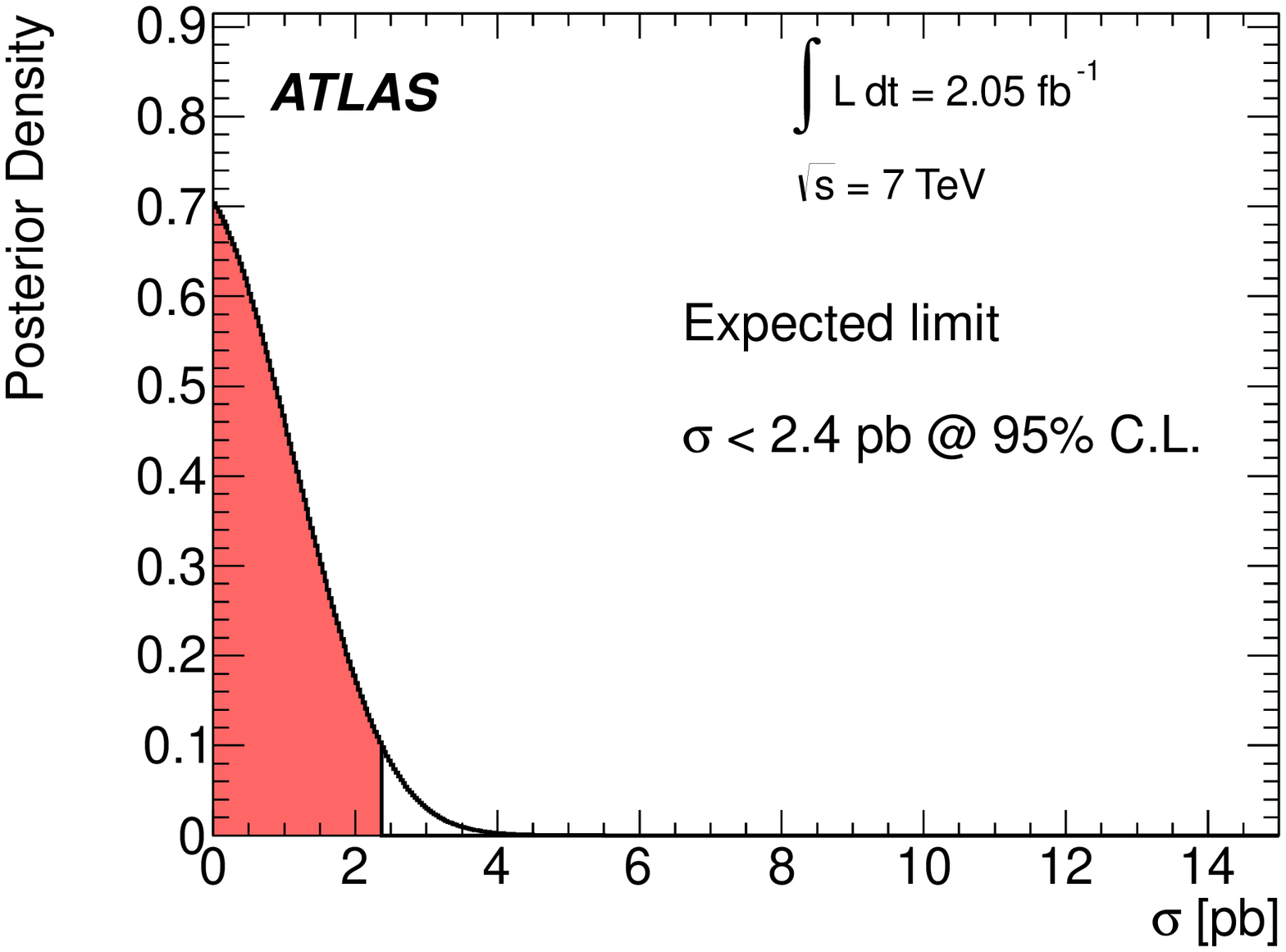}\\[-1em]
(a) \hspace*{0.5\textwidth} \\
\includegraphics[width=0.45\textwidth]{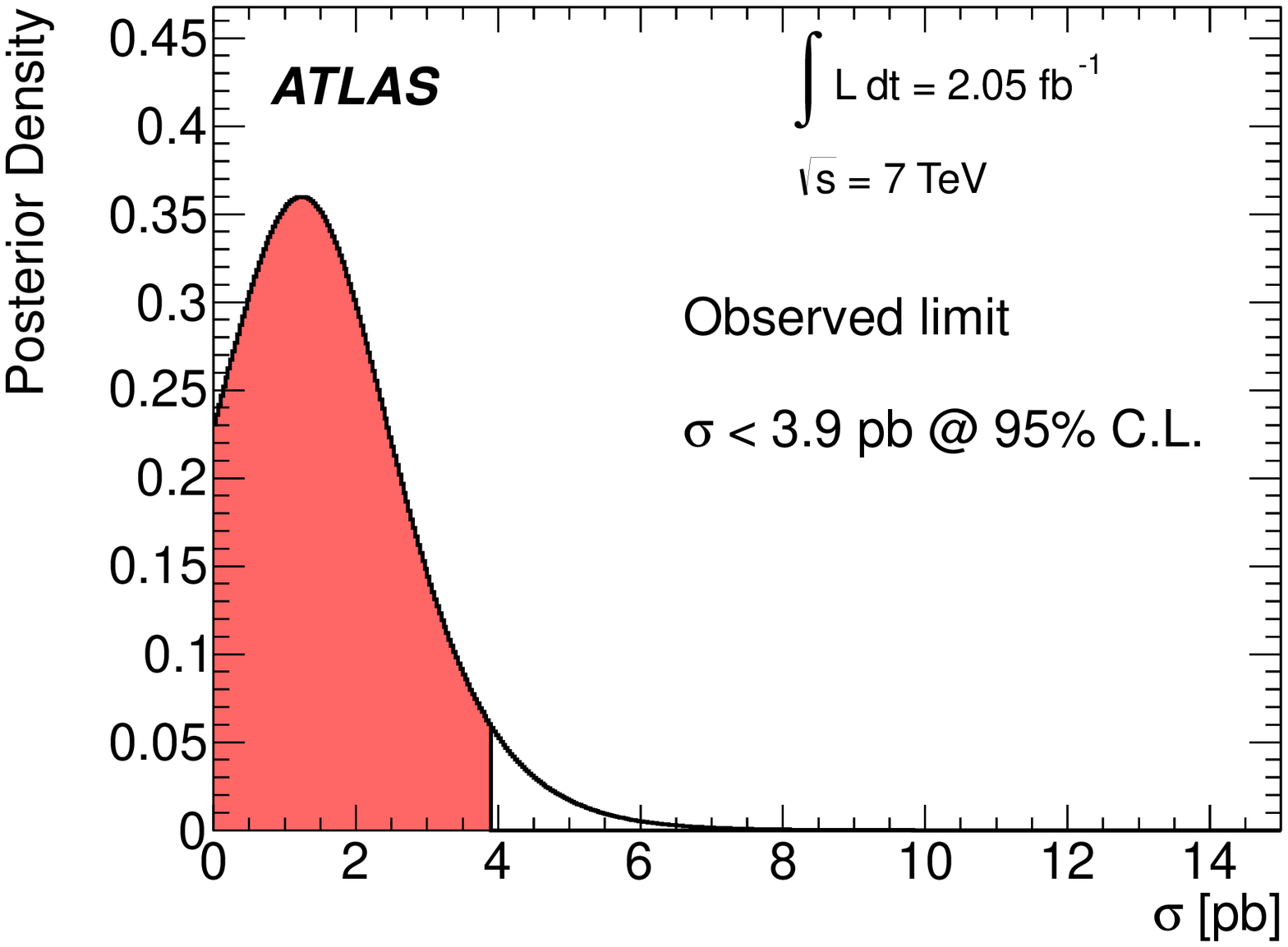}\\[-1em]
(b) \hspace*{0.5\textwidth} \\
\caption{\label{fig:final_pdf_couplings}
Distribution of the posterior probability function including all systematic uncertainties for 
(a) the expected upper limit and  (b) the observed upper limit at 95\% C.L..}
\end{figure}

\section{Results}
\label{sec:result}

A Bayesian statistical analysis~\cite{Bertram:2000br,jaynes} using a binned likelihood method applied to 
the neural network output distributions for the electron and muon channel combined is performed to measure or set 
an upper limit on the FCNC single top-quark production cross-section.

\begin{figure}[!h]
\centering
\includegraphics[width=0.45\textwidth]{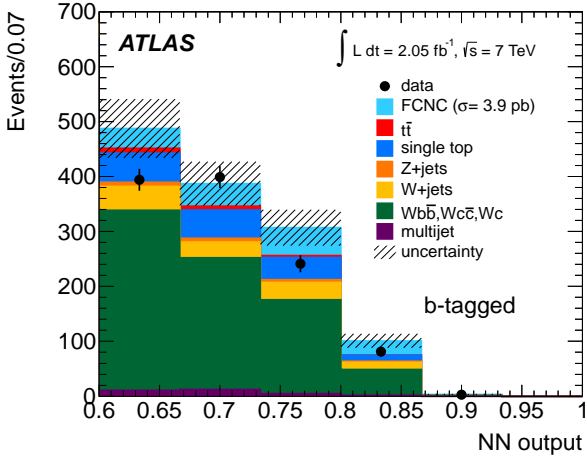}\\
\caption{\label{fig:finalnn}
Distributions of the neural network output: Observed signal and simulated background output distribution normalised to the mean value of the marginalised 
nuisance parameters, zoomed into the signal region. The FCNC single top quark process is normalised to the observed limit of 3.9~pb. 
The hatched band indicates the statistical uncertainty from the sizes of the simulated samples and the uncertainty in the background normalisation.}
\end{figure}

\begin{figure}[!h]
\centering
\includegraphics[width=0.45\textwidth]{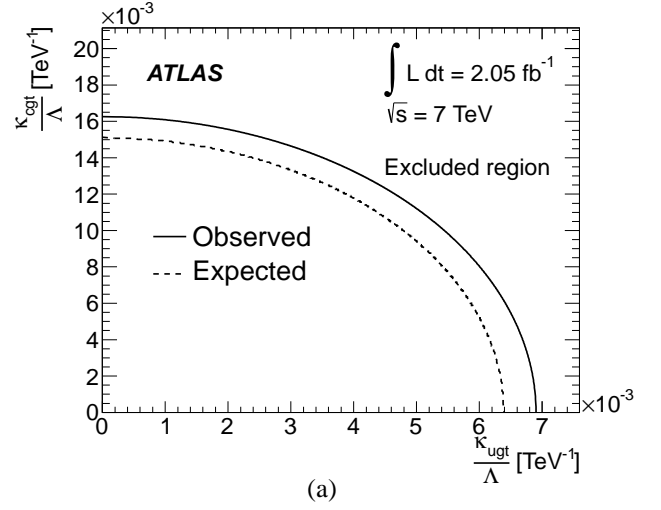}\\[-1em]
(a) \hspace*{0.5\textwidth} \\
\includegraphics[width=0.45\textwidth]{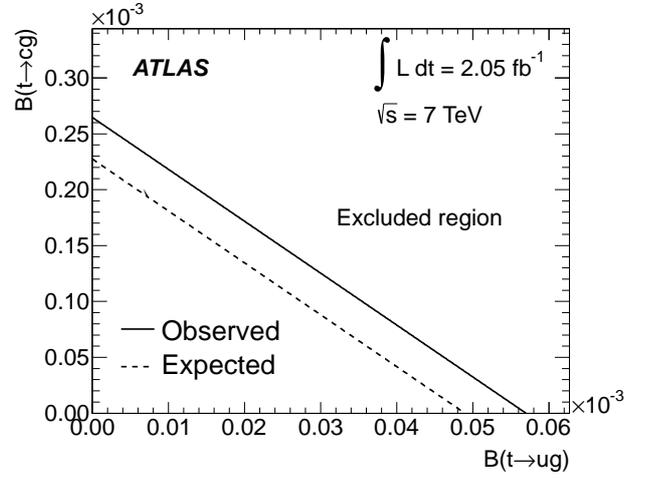}\\[-1em]
(b) \hspace*{0.5\textwidth} \\
\caption{\label{fig:limit}
Upper limit (a) on the coupling constants $\kappa_{ugt}/\Lambda$ and $\kappa_{cgt}/\Lambda$ and (b) on the branching fractions $t \rightarrow ug$ and $t \rightarrow cg$.}
\end{figure}

Systematic uncertainties and their correlations among processes are included with a direct sampling approach where the same Gaussian shift
is applied to each source, process, and bin for a given uncertainty.
The posterior density function (pdf) is obtained by creating a large number
of samples of systematic shifts. A separate likelihood distribution is obtained for each sample, and the final
pdf is then the average over all of the individual likelihoods.
This pdf gives the probability of the signal hypothesis as a function of the signal cross-section.
Since no significant rate of FCNC single top-quark production is observed, an upper limit is set by integrating the pdf.
To estimate the a priori sensitivity, we use a pseudo-dataset corresponding to the prediction from simulations (Asimov dataset)~\cite{Cowan:2010js}
and treated in the same way as the observed dataset.
The resulting expected upper limit at 95\% confidence level (C.L.)  on the anomalous FCNC single top-quark production cross-section 
including all systematic uncertainties is 2.4 pb, while the corresponding observed upper limit is 3.9~pb,
as shown in Figures~\ref{fig:final_pdf_couplings}(a) and \ref{fig:final_pdf_couplings}(b), respectively.
To visualise the observed upper limit in the neural network output distribution 
Figure~\ref{fig:finalnn} shows the FCNC single top-quark process scaled to observed upper limit on top of the SM background processes.
As a cross-check we performed the full statistical analysis only for events with NN output $>0$, which yields an observed upper limit at 95\% C.L. of 5.9 pb.
Using the NLO predictions for the FCNC single top-quark production cross-section~\cite{Gao:2011fx,Liu:2005dp}, the measured upper limit on the production cross-section is converted into 
limits on the coupling constants $\kappa_{ugt}/\Lambda$ and $\kappa_{cgt}/\Lambda$. Assuming $\kappa_{cgt}/\Lambda = 0$ one finds $\kappa_{ugt}/\Lambda < 6.9\cdot 10^{-3}$~TeV$^{-1}$ and assuming $\kappa_{ugt}/\Lambda = 0$ one finds $\kappa_{cgt}/\Lambda < 1.6\cdot 10^{-2}$~TeV$^{-1}$. Figure~\ref{fig:limit}(a) shows the distribution of the upper limit for all possible combinations.  
Using the NLO calculation~\cite{Zhang:2008yn}, upper limits on the branching fractions ${\cal B}(t \rightarrow ug) < 5.7\cdot 10^{-5}$ assuming ${\cal B}(t \rightarrow cg)=0$, and  ${\cal B}(t \rightarrow cg) < 2.7\cdot 10^{-4}$ assuming ${\cal B}(t \rightarrow ug)=0$ are derived, as shown in Figure~\ref{fig:limit}(b).

\section{Conclusion}
In summary, a data sample selected to consist of events with an isolated electron or muon, missing transverse momentum and a $b$-quark jet 
has been used to search for FCNC production of single top-quarks at the LHC.
No evidence for such processes is found and the upper limit at 95\% C.L. on the production cross-section is 3.9 pb.
The limits set on the coupling constants $\kappa_{ugt}/\Lambda$ and $\kappa_{cgt}/\Lambda$ 
and the branching fractions ${\cal B}(t \rightarrow ug) < 5.7\cdot 10^{-5}$ assuming ${\cal B}(t \rightarrow cg)=0$, 
and  ${\cal B}(t \rightarrow cg) < 2.7\cdot 10^{-4}$ assuming ${\cal B}(t \rightarrow ug)=0$
are the most stringent to date
on FCNC single top-quark production processes for $qg \rightarrow t$ and improve on the previous best limits~\cite{Abazov:2010qk}
by factors of 4 and 15, respectively.

%


\section{Acknowledgements}

We thank CERN for the very successful operation of the LHC, as well as the
support staff from our institutions without whom ATLAS could not be
operated efficiently.

We acknowledge the support of ANPCyT, Argentina; YerPhI, Armenia; ARC,
Australia; BMWF, Austria; ANAS, Azerbaijan; SSTC, Belarus; CNPq and FAPESP,
Brazil; NSERC, NRC and CFI, Canada; CERN; CONICYT, Chile; CAS, MOST and NSFC,
China; COLCIENCIAS, Colombia; MSMT CR, MPO CR and VSC CR, Czech Republic;
DNRF, DNSRC and Lundbeck Foundation, Denmark; EPLANET and ERC, European Union;
IN2P3-CNRS, CEA-DSM/IRFU, France; GNAS, Georgia; BMBF, DFG, HGF, MPG and AvH
Foundation, Germany; GSRT, Greece; ISF, MINERVA, GIF, DIP and Benoziyo Center,
Israel; INFN, Italy; MEXT and JSPS, Japan; CNRST, Morocco; FOM and NWO,
Netherlands; RCN, Norway; MNiSW, Poland; GRICES and FCT, Portugal; MERYS
(MECTS), Romania; MES of Russia and ROSATOM, Russian Federation; JINR; MSTD,
Serbia; MSSR, Slovakia; ARRS and MVZT, Slovenia; DST/NRF, South Africa;
MICINN, Spain; SRC and Wallenberg Foundation, Sweden; SER, SNSF and Cantons of
Bern and Geneva, Switzerland; NSC, Taiwan; TAEK, Turkey; STFC, the Royal
Society and Leverhulme Trust, United Kingdom; DOE and NSF, United States of
America.

The crucial computing support from all WLCG partners is acknowledged
gratefully, in particular from CERN and the ATLAS Tier-1 facilities at
TRIUMF (Canada), NDGF (Denmark, Norway, Sweden), CC-IN2P3 (France),
KIT/GridKA (Germany), INFN-CNAF (Italy), NL-T1 (Netherlands), PIC (Spain),
ASGC (Taiwan), RAL (UK) and BNL (USA) and in the Tier-2 facilities
worldwide.




\bibliographystyle{model1a-num-names}
\bibliography{fcnc_singletop}







\clearpage
\onecolumn
\begin{flushleft}
{\Large The ATLAS Collaboration}

\bigskip

G.~Aad$^{\rm 48}$,
B.~Abbott$^{\rm 110}$,
J.~Abdallah$^{\rm 11}$,
A.A.~Abdelalim$^{\rm 49}$,
A.~Abdesselam$^{\rm 117}$,
O.~Abdinov$^{\rm 10}$,
B.~Abi$^{\rm 111}$,
M.~Abolins$^{\rm 87}$,
O.S.~AbouZeid$^{\rm 157}$,
H.~Abramowicz$^{\rm 152}$
% ...
\bigskip

% ...
$^{10}$ Institute of Physics, Azerbaijan Academy of Sciences, Baku, Azerbaijan\\
$^{11}$ Institut de F\'isica d'Altes Energies and Departament de F\'isica de la Universitat Aut\`onoma  de Barcelona and ICREA, Barcelona, Spain\\
% ...
$^{48}$ Fakult\"{a}t f\"{u}r Mathematik und Physik, Albert-Ludwigs-Universit\"{a}t, Freiburg i.Br., Germany\\
$^{49}$ Section de Physique, Universit\'e de Gen\`eve, Geneva, Switzerland\\
% ...
$^{87}$ Department of Physics and Astronomy, Michigan State University, East Lansing MI, United States of America\\
% ...
$^{110}$ Homer L. Dodge Department of Physics and Astronomy, University of Oklahoma, Norman OK, United States of America\\
$^{111}$ Department of Physics, Oklahoma State University, Stillwater OK, United States of America\\
% ...
$^{117}$ Department of Physics, Oxford University, Oxford, United Kingdom\\
% ...
$^{152}$ Raymond and Beverly Sackler School of Physics and Astronomy, Tel Aviv University, Tel Aviv, Israel\\
% ...
$^{157}$ Department of Physics, University of Toronto, Toronto ON, Canada\\
% ...
\end{flushleft}


\end{document}